\documentclass[12pt]{article}
\usepackage{amssymb}
\def\beq{\begin{equation}}
\def\eeq{\end{equation}}
\usepackage[all,2cell,dvips]{xy}
\newtheorem{proposicion}{Proposition}
\newtheorem{demosprop}{Proof of Proposition}
\newtheorem{teorema}{Theorem}
\newtheorem{demosteor}{Proof of Theorem}
\usepackage[all,2cell,dvips]{xy}
\def\IR{\relax{\rm I\kern -.18em R}}
\begin{document}
\title{The Gardner Category and Non-local Conservation Laws for $N=1$ Super KdV }
\author{ \Large S. Andrea*, A. Restuccia**, A. Sotomayor***}
\maketitle{\centerline {*Departamento de Matem\'{a}ticas,}}
\maketitle{\centerline{**Departamento de F\'{\i}sica}}
\maketitle{\centerline{Universidad Sim\'on Bol\'{\i}var}}
\maketitle{\centerline{***Departamento de Ciencias B\'{a}sicas}}
\maketitle{\centerline{Unexpo, Luis Caballero Mej\'{\i}as }}
\maketitle{\centerline{e-mail: sandrea@usb.ve, arestu@usb.ve,
sotomayo@fis.usb.ve }}
\begin{abstract} The non-local conserved quantities of $N=1$ Super
KdV are obtained using a complete algebraic framework where the
Gardner category is introduced. A fermionic substitution semigroup
and the resulting Gardner category are defined and several
propositions concerning their algebraic structure are proven. This
algebraic framework allows to define general transformations
between different nonlinear SUSY differential equations. We then
introduce a SUSY ring extension to deal with the non-local
conserved quantities of SKdV. The algebraic version of the
non-local conserved quantities is solved in terms of the
exponential function applied to the $D^{-1}$ of the local
conserved quantities of SKdV. Finally the same formulas are shown
to work for rapidly decreasing superfields.

\end{abstract}
\section{Introduction} The supersymmetric algebra is the unique
extension of the super-Poincar\'e algebra which is consistent with
the $S$-matrix of quantum field theory. The most remarkable SUSY
theory explains how superstrings and other extended SUSY objects
can be consistently tied together in what also has been called
$M$-theory.

Free (string) superstring theory ia a two-dimensional
supersymmetric theory whose local symmetry group is generated by
the (Virasoro) Super-Virasoro algebra.

These algebras may be realized as algebras of the (potential)
superpotential of (KdV) SKdV  \cite{M1,Manin} equations when the
second Hamiltonian structure (with the corresponding Poisson
structure) is considered \cite{M2}.

It is then reasonable to think that the hierarchy of (KdV) SKdV is
related to the loop expansion of (string) superstring theory in
terms of the genus of Riemann surfaces \cite{Witten}.

The SKdV hierarchy also arises from supersymmetric quantum
mechanics. In fact, it was proven in \cite{Andrea2, Andrea3} that
the entire SKdV hierarchy appears in the asymptotic expansion of
the Green's function $g(x,\theta,y,\theta^\prime)$ of the super
heat operator, as $t\rightarrow0^+$ and
$g(x,\theta,y,\theta^\prime)$ is restricted to the diagonal
$x=y,\theta=\theta^\prime.$ The same result holds for the pure
``bosonic" (non-SUSY) KdV hierarchy arising from the Green's
function of the heat operator with potential, that is, the
``euclidean" Schr\"{o}dinger operator \cite{Avramidi}.

The KdV equation has an infinite number of discrete conserved
quantities (CQ). The SUSY extension of these conserved quantities
are also CQ for the SKdV equation; but a remarkable difference
between the two equations is that SKdV has a second sequence of
CQ, these being non-local and intrinsically supersymmetric in
nature. They have been interpreted \cite{M3} as the Poisson square
root of the local CQ's, in the sense that
\[\{J,J^\prime\}=H
\] where $J$ and $J^\prime$ are non-local CQ's and $H$ is a local
CQ of SKdV.

The conservation laws of KdV and SKdV may be obtained from the Lax
representations of these equations; for a review see \cite{M3}.
The non-local CQ of SKdV were first obtained by analyzing the
infinite set of symmetries of SKdV, eg. \cite{Kersten}. Later on
they were obtained from the Lax operator in \cite{M4 and Dargis}.

Another way to obtain these conservation laws is trough the
supersymmetric extension \cite{M1, Andrea1} of Gardner
transformation \cite{Gardner et al}. It may be interpreted as a
one-parameter integrable deformation of SKdV. The deformation is
\[\phi=\chi+\varepsilon D^2\chi-\varepsilon^2\chi D\chi,\] where
$\varepsilon$ is the deformation parameter.

If the superfield $\chi$ satisfies the $S$-Gardner equation
\cite{M1} then $\phi$ satisfies SKdV. Then, using the fact that
$H=\int dxd\theta\chi$ is a conserved quantitie of the $S$-Gardner
equation, it was shown \cite{M1} that all the local conserved
quantities of SKdV arise in the formal expansion of $H$ in powers
of $\varepsilon$.

It was left as an open problem, OP1 in the review of P. Mathieu
\cite{M3}, to find the non-local conserved quantities of SKdV from
some integrable $\varepsilon$-deformation.

In the present paper OP1 is solved, by first rephrasing it in a
completely algebraic framework working first in the free SUSY
derivation ring constructed in \cite{Andrea2}, a fermionic
substitution semigroup is introduced. The resulting Gardner
category is an algebraic construction modelled on the possibility
of more general Gardner transforms between different nonlinear
SUSY differential equations. In the particular case of SKdV the
local conserved quantities are constructed from this formalism.

We then introduce SUSY ring extensions in order to deal with the
possibility of non-local conserved quantities. The algebraic
version of the non-local CQ problem is solved, using the
exponential function applied to the $D^{-1}$ of the local
conserved quantities which the ring extensions provide.

Finally the same formulas are shown to work for rapidly decreasing
superfields, and the non-local CQ's so obtained are shown to agree
with some found in the literature.

\section{The Fermionic Substitution Semigroup  } Let $
\mathcal{A}$ be the free SUSY derivation ring on a single
fermionic generator $a_1$. This ring is generated by its fermionic
elements $a_1,a_3, a_5,\ldots$ and bosonic elements $a_2,a_4,
a_6,\ldots$ and its superderivation $ D:\mathcal{A}\rightarrow
\mathcal{A}$ is determinated by $Da_n=a_{n+1}$ for $n\geq1$.

The ring extension $ \mathcal{A}[\varepsilon]\supset\mathcal{A}$
consists of all formal power series
$h=\sum_{n=0}^\infty\varepsilon^nh_n$ with coefficients
$h_n(a_1,a_2,\ldots)\in\mathcal{A}.$ Its involution
$h\rightarrow\bar{h}$ and superderivation $h\rightarrow Dh$ are
defined componentwise from the same operations in $\mathcal{A}$.
The supercommutativity equation $gh=\pm hg$ holds when
$\bar{g}=\pm g$ and $ \bar{h}=\pm h$, a minus sign when
$\bar{g}=-g$ and $ \bar{h}=-h$, and a plus sign in the other three
cases.

When $f,g\in\mathcal{A}[\varepsilon]$ with $ \bar{f}=-f$, the
substitution of $f$ in $g$ produces another element $g\circ
f\in\mathcal{A}[\varepsilon]$. It is defined by the formulas
\begin{eqnarray*}& & g=\sum_0^\infty\varepsilon^ng_n\left(a_1,a_2,\ldots
\right)\\ & & g\circ
f=\sum_0^\infty\varepsilon^ng_n\left(f,Df,\ldots \right)
   \end{eqnarray*} The following properties are derived.
\begin{proposicion} When $f,g,h\in \mathcal{A}[\varepsilon]$ with
$\bar{f}=-f$ one has \begin{eqnarray*}& & \left(g+h \right)\circ f=\left(g\circ f \right)+\left(h\circ f\right)\\
& & \left(gh \right)\circ f=\left(g\circ f \right)\left(h\circ f
\right),
   \end{eqnarray*} which is to say that the operation
   $g\rightarrow g\circ f$ is a ring homomorphism $
   \mathcal{A}[\varepsilon]\rightarrow \mathcal{A}[\varepsilon],$
   for any fixed $ \bar{f}=-f.$

\end{proposicion}
\begin{demosprop}{\em It suffices to take $g,h\in \mathcal{A}.$
Since $\mathcal{A}[\varepsilon]$ is supercommutative and $f$ is
fermionic, there is no ambiguity in passing from
$g(a_1,a_2,\ldots)h(a_1,a_2,\ldots)$ to
$g(f,Df,\ldots)h(f,Df,\ldots).$
}

\end{demosprop}

$\smallskip$

\begin{proposicion} When $f,g\in \mathcal{A}[\varepsilon]$ with $
\bar{f}=-f$, one has \[D\left( g\circ f\right)=\left(Dg
\right)\circ f.\]

\end{proposicion}

\begin{demosprop}{\em When $g\in \mathcal{A}$ and is just some
$a_n$, both sides of the equation give $D^nf.$

Suppose now that the proposition is true for some
$g,h\in\mathcal{A}.$ Then when the Proposition 1 is applied to the
equation \[D\left( gh\right)=\left(Dg \right)h+\bar{g}\left(Dh
\right)
\] we obtain
\[\left(D(gh)  \right)\circ f=\left((Dg)\circ f \right)\left(h\circ
f  \right)+\left(\bar{g}\circ f \right) \left((Dh)\circ f
\right).   \] On the other hand
\begin{eqnarray*}D\left(gh\circ f \right)&=& D\left(\left(g\circ f
\right)\left(h\circ f \right) \right) \\ & = & \left(\left(Dg\circ
f \right) \right)\left(h\circ f \right)+\overline{\left(g\circ f
\right)}D\left(h\circ f \right).\end{eqnarray*}

Since $\overline{\left(g\circ f \right)}=\overline{g}\circ f,$ the
desired equality for $gh$ follows from its truth for $g$ and $h$.
It follows that Proposition 2 holds for any element $g$ of $
\mathcal{A}$, and hence also for $g\in \mathcal{A}[\varepsilon]$.

}

\end{demosprop}

$\bigskip$

Now let $
\mathcal{A}_1[\varepsilon]\subset\mathcal{A}[\varepsilon]$ be the
subset of all fermionic elements. The substitution product gives
$g\circ f\in \mathcal{A}_1[\varepsilon]$ if
$g,h\in\mathcal{A}_1[\varepsilon].$

\begin{proposicion} The substitution product is associative:
\[\left(h\circ g \right)\circ f=h\circ\left(g\circ f \right).
 \]

\end{proposicion}
\begin{demosprop}{\em It suffices to treat the case
\[h\left(a_1,a_2,\ldots\right)\in\mathcal{A}_1[\varepsilon]\supset\mathcal{A}.\]
By definition \[h\circ \left(g\circ f \right) =h\left(\left(g\circ
f \right),D\left(g\circ f \right),\ldots \right).\] But from
Proposition 2 \[h\circ \left(g\circ f \right)=h\left(\left(g\circ
f \right),\left(Dg \right)\circ f,\ldots \right)
\] But $h$ is just a sum of products of $a_1,a_2,\ldots$ and the
$f$-substitution is a ring homomorphism.

Therefore \[h\circ \left(g\circ f \right)=h\left(g,Dg,\ldots
\right)\circ f=\left(h\circ g \right)\circ f, \] and the proof is
complete.

}\end{demosprop}

$\smallskip$

Thus $ \mathcal{A}_1[\varepsilon]$ is made into a semigroup by the
substitution construction. Evidently the element
$a_1\in\mathcal{A}_1\subset\mathcal{A}_1[\varepsilon]$ acts as the
identity element of this semigroup.

When an element of $\mathcal{A}_1[\varepsilon]$ has the value
$a_1$ when $\varepsilon=0$ it is invertible:

\begin{proposicion} Given $f=a_1+\varepsilon
f_1+\varepsilon^2f_2+\cdots\in\mathcal{A}_1[\varepsilon]$ there
exists $g=a_1+\varepsilon
g_1+\varepsilon^2g_2+\cdots\in\mathcal{A}_1[\varepsilon]$ with
$g\circ f=a_1$.

\end{proposicion}
\begin{demosprop}{\em For any $h(a_1,a_2,\ldots)\in\mathcal{A}$
and $f$ as above, $h\circ f=h+\sum_{k=1}^\infty\varepsilon^kh_k$
for some $h_k\in\mathcal{A}$. Therefore\[\left( a_1-\varepsilon
f_1\right)\circ f=\sum_{k=2}^\infty\varepsilon^k\tilde{h_k} \] for
some $\tilde{h_k}\in\mathcal{A}.$

If $g_1,\ldots g_n\in\mathcal{A}$ have been found such that
\[\left(a_1+\sum_{k=1}^n\varepsilon^kg_k \right)\circ
f=\varepsilon^{n+1}r+\cdots
\] then the choice $g_{n+1}=-r\in\mathcal{A}$ gives the same
equation, but for $n+1$. Thus all the coefficients of
$g=a_1+\sum_{k+1}^\infty\varepsilon^kg_k$ are determined
recursively, and the proof is complete.

}\end{demosprop}

$\smallskip$

An easy corollary shows that left and right inverses are the same.
\begin{proposicion} Given
$f=a_1+\sum_{k=1}^\infty\varepsilon^kf_k$ and
$g=a_1+\sum_{k=1}^\infty\varepsilon^kg_k$ in $
\mathcal{A}_1[\varepsilon]$. If $f\circ g=a_1$ then $g\circ
f=a_1$.

\end{proposicion}
\begin{demosprop}{\em There exists $h$ with $h\circ f=a_1.$ Then
$(h\circ f)\circ g=a_1\circ g=g$ while $h\circ(f\circ g)=h\circ
a_1=h.$ Thus $h=g$ and $h\circ f=g\circ f=a_1$, completing the
proof.}\end{demosprop}

$\smallskip$

As an exercise one can compute the inverse of $f=a_1+\varepsilon
a_1a_2$, obtaining $g=a_1-\varepsilon
a_1a_2+2\varepsilon^2a_1a_2^2-5\varepsilon^3a_1a_2^3+\cdots,$ the
coefficient of $\varepsilon^n$ being ${(-1)}^n{2n+2\choose
n+1}\frac{a_1a_2^n}{4n+2}.$

$\smallskip$

\subsection{Frechet Derivative Operator}
Associated with the ring $ \mathcal{A}$ and its superderivation
$D:\mathcal{A}\rightarrow\mathcal{A}$ there is a ring $
\mathcal{O}_p\mathcal{A}$ whose elements are the finite order
differential operators $L=\sum_{k=0}^Nl_kD^k$ with
$l_k\in\mathcal{A}.$ Each $L$ acts linearly in $ \mathcal{A}$, and
the product of two operators is computed from repeated
applications of the SUSY product rule $D(gh)=(Dg)h+\bar{g}(Dh).$

When $ \bar{f}=-f\in\mathcal{A}_1[\varepsilon],$ the substitution
of $f$ in $L$ is defined by \[L\circ f=\sum_{k=0}^N\left(l_k\circ
f \right)D^k.
\] Thus $L\circ f,$ a formal power series with operator
coefficients, is in the ring $
(\mathcal{O}_p\mathcal{A})[\varepsilon]$ whose elements are the
sums $\sum_{m,n}\varepsilon^ml_{m,n}D^n$ with
$l_{m,n}\in\mathcal{A}$ and $l_{m,n}=0$ for $n>>0$, at any given
$m$.

Given $L\in(\mathcal{O}_p\mathcal{A})[\varepsilon]$ and
$h\in\mathcal{A}[\varepsilon]$, the element $Lh\in
\mathcal{A}[\varepsilon]$ is well-defined because
$(\sum^m\varepsilon^mL_m)(\sum\varepsilon^nh_n)$ is again a power
series with coefficients in $ \mathcal{A}$. The effect of
$f$-substitution is as to be expected.

$\smallskip$

\begin{proposicion} If
$L\in(\mathcal{O}_p\mathcal{A})[\varepsilon]$,
$h\in\mathcal{A}[\varepsilon]$, and
$f\in\mathcal{A}_1[\varepsilon],$ then
\[\left(Lh  \right)\circ f=\left(L\circ f  \right)\left( h\circ f \right). \]
\end{proposicion}
\begin{demosprop}{\em When $l,h\in\mathcal{A}$ one has \[\left(lD^nh\right)\circ f=\left(l\circ f \right)
D^n\left(h\circ f  \right)\] by Propositions 1 and 2. The general
case reduces to linear combinations of this special
case.}\end{demosprop}

$\smallskip$

The foregoing constructions come into play when we ask for the
first variation of the substitution operation. Given any
$f=\sum_{m=0}^\infty\varepsilon^mf_m\left(a_1,a_2,\ldots \right)$
in $ \mathcal{A}_1[\varepsilon]$, its Frechet derivative operator
is
\[f^\prime=\sum_{m=0}^\infty\sum_{n=1}^\infty\varepsilon^m\frac{\partial}{\partial
a_n}f_m\left(a_1,a_2,\ldots
\right)D^{n-1}\in(\mathcal{O}_p\mathcal{A})[\varepsilon].
\] Then for any $\psi\in\mathcal{A}_1[\varepsilon]$ the
substitution by $a_1+t\psi\in\mathcal{A}_1[\varepsilon]$ gives
\[f\circ\left(a_1+t\psi  \right)=f+tf^\prime\psi+\cdots,  \] the
full right side of the equation being a power series in $t$ with $
\mathcal{A}[\varepsilon]$ coefficients.

A more general formula appears when $a_1$ is replaced by
$\varphi\in\mathcal{A}_1[\varepsilon]$ and
\[f\circ\left(\varphi+t\psi  \right)=f\circ\varphi+t\left(f^\prime\circ\varphi \right)\psi+\cdots,   \]
valid when $f,\varphi,\psi\in\mathcal{A}_1[\varepsilon].$

The chain rule is now immediate.

\begin{proposicion} When $f,g\in\mathcal{A}_1[\varepsilon]$ one
has \[{\left( f\circ g  \right)}^\prime=\left(f^\prime\circ g
\right)g^\prime.\]
\end{proposicion}
\begin{demosprop}{\em From the definition,
${(f\circ g)}^\prime\in(\mathcal{O}_p\mathcal{A})[\varepsilon]$ is
given by \[f\circ g\circ\left(a_1+t\psi \right)=f\circ g+t{(f\circ
g)}^\prime\psi+\cdots
\] where $\psi\in\mathcal{A}_1[\varepsilon]$ is arbitrary. But
\[g\circ\left(a_1+t\psi \right)=g+tg^\prime\psi+\cdots,  \] giving
\begin{eqnarray*}f\circ g\circ\left(a_1+t\psi  \right)&=&
f\circ\left(g+tg^\prime\psi+\cdots  \right) \\ &=& f\circ
g+t\left(f^\prime\circ g \right)g^\prime\psi+\cdots\end{eqnarray*}
Since $\psi\in\mathcal{A}_1[\varepsilon]$ is arbitrary, the proof
is complete.

}\end{demosprop}

\section{The Gardner Category} An element
$f=\sum_{m=0}^\infty\varepsilon^mf_m(a_1,a_2,\ldots)$ of $
\mathcal{A}_1[\varepsilon]$ may be taken to represent a nonlinear
differential equation \[\frac{\partial}{\partial
t}\alpha(x,t)=\sum_{m=0}^\infty\varepsilon^mf_m\left(\alpha\left(x,t
\right),D\alpha\left(x,t \right),\ldots \right) \] if
$\alpha(x,t)$ is a fermionic superfield and the superderivation
$D_1=\frac{\partial}{\partial
\theta}+\theta\frac{\partial}{\partial x}$ is also known as the
covariant derivative.

A second element $g\in\mathcal{A}_1[\varepsilon]$ represents a
second differential equation, for an unknown superfield
$\beta(x,t)$.

Then given a third element $r\in\mathcal{A}_1[\varepsilon]$, one
might want the
transformation\[\beta(x,t)=\sum_{m=0}^\infty\varepsilon^m
r_m\left(\alpha\left(x,t \right),D\alpha\left(x,t \right),\ldots
\right)
\] to transform solutions of the first equation into solutions of
the second. After some computation one sees that this happens if
\[g\circ r=r^\prime f.\]

Accordingly, $f$ and $g$ can be called ``objects" in the Gardner
category, and $r$ a ``morphism" from $f$ to $g$, written
$\xymatrix{\  g & \ar[l]_r  f,
 }$ if the above equality holds in $ \mathcal{A}_1[\varepsilon].$

 Obviously the choice $r=a_1\in\mathcal{A},$
 $r^\prime=I\in\mathcal{O}_p\mathcal{A}$ gives the identity
 automorphism of each object.

 But the composition of morphisms must be checked.

 \begin{proposicion} Given
 $f,g,h,r,s\in\mathcal{A}_1[\varepsilon].$

 If $\xymatrix{\  h & \ar[l]_s  g
 }$ and $\xymatrix{\  g & \ar[l]_r  f
 }$ then $\xymatrix{\  h & \ar[l]_{s\circ r}  f.
 }$

 \end{proposicion}
 \begin{demosprop}{\em From $h\circ s=s^\prime g$ it follows that
 \begin{eqnarray*}h\circ s\circ r &=& s^\prime g\circ r \\ &=&\left(s^\prime\circ r \right)\left(
 g\circ r\right)    \end{eqnarray*} after applying Proposition
 6 to $s^\prime\in\mathcal{O}_p\mathcal{A}[\varepsilon]$ and
 $g\in\mathcal{A}_1[\varepsilon].$ But $g\circ r=r^\prime f,$
 giving \begin{eqnarray*}h\circ\left(s\circ r \right) &=& \left( s^\prime\circ r\right)r^\prime f
 \\ &=& {\left(s \circ r\right)}^\prime f   \end{eqnarray*} by the
 chain rule, Proposition 7. This completes the proof.
 }\end{demosprop}

 $\smallskip$

 The possibility of isomomorphism classes in the Gardner category
 leads one to examine the invertible elements.
 \begin{proposicion} Given $f,g,r,s\in\mathcal{A}_1[\varepsilon]$
 with $r\circ s=s\circ r=a_1.$

 If  $\xymatrix{\  g & \ar[l]_r  f
 }$

 then  $\xymatrix{\  f & \ar[l]_s  g.
 }$

 \end{proposicion}
 \begin{demosprop}{\em From $g\circ r\circ s=r^\prime f\circ s$
 one obtains \begin{eqnarray*}& & g=\left(r^\prime \circ s
 \right)\left(f\circ s  \right) \\ & & s^\prime
 g=s^\prime\left(r^\prime\circ s
 \right)\left(f\circ s  \right).
 \end{eqnarray*} The desired conclusion $f\circ s^\prime=s^\prime
 g$ would follow from $s^\prime(r^\prime\circ s)=I.$ But the
 $s$-substitution is also a homomorphism of the ring of operators,
 and it converts the known $(s^\prime\circ r)r^\prime=I$ into
 $s^\prime(r^\prime\circ s)=I.$ This completes the proof.

 }\end{demosprop}

 \subsection{The Gardner Transform}
 It is known that \[\xymatrix{\  g & \ar[l]_r  f
 }\] where $g=a_7+3a_1a_4+3a_2a_3\in\mathcal{A}$ represents the
 SUSY KdV equation, $r=a_1+\varepsilon
 a_3-\varepsilon^2a_1a_2\in\mathcal{A}[\varepsilon]$ represents
 the Gardner transform, and $f$ is a certain modification of the
 KdV equation.

 In general when $h\in\mathcal{A}$ one has $h\circ r=h+\varepsilon
 D^2h+\cdots$ because $D^2:\mathcal{A}\rightarrow\mathcal{A}$ is
 an ordinary derivation. Since \[r^\prime=I+\varepsilon D^2-\varepsilon^2\left(a_2I+a_1D \right), \]
 the difference $h\circ r-r^\prime h$ will always have the form
 $\sum_{n=2}^\infty\varepsilon^nh_n.$

 To compute this difference for $h=a_7$ we must substract
 $-(a_2+a_1D)a_7)$ from $-D^6(a_1a_2),$ obtaining
 \[a_7\circ r-r^\prime a_7=-3\varepsilon^2\left(a_3a_6+a_4a_5  \right). \]
 For the second term $h=a_1a_2$ we note first that
 $(a_2I+a_1D)a_1a_2=2a_1a_2^2,$ giving
 \[\left(a_1a_2\right)\circ r=rDr=\cdots+\varepsilon^2\left(a_3a_4-2a_1a_2^2  \right)-\varepsilon^3\left(
 a_1a_2a_4+a_2^2a_3 \right)+\varepsilon^4\left(a_1a_2^3 \right).
 \]This gives \[\left(a_1a_2\right)\circ r-r^\prime\left(a_1a_2 \right)=\varepsilon^2a_3a_4-\varepsilon^3\chi
 +\varepsilon^4\varrho \] with
 $\chi=a_1a_2a_4+a_2^2a_3,\varrho=a_1a_2^3.$ These elements of $
 \mathcal{A}$ satisfy $(a_2I+a_1D)\chi=D^2\varrho$ and
 $(a_4I+a_3D)(a_1a_2)=\chi.$ Thus
 \[\varepsilon^2r^\prime \chi=\varepsilon^2\chi+\varepsilon^3D^2\chi-\varepsilon^4D^2\varrho, \]
 and \[D^2\left(\left(a_1a_2  \right)\circ r-r^\prime\left(a_1a_2  \right)    \right)=
 \varepsilon^2\left(a_3a_6+a_4a_5+\chi-r^\prime\chi  \right). \]
 In order to pass to $D^2(a_1a_2)=a_1a_4+a_2a_3,$ we note the
 operator commutator equation \[D^2r^\prime=r^\prime D^2-\varepsilon^2\left(a_4I+a_3D  \right), \]
 which gives
 \[D^3r^\prime\left(a_1a_2  \right)=r^\prime\left(a_1a_4+a_2a_3  \right)-\varepsilon^2\chi.  \]
 Together with Proposition 2 this gives
 \[\left( a_1a_4+a_2a_3 \right)\circ r-r^\prime\left(a_1a_4+a_2a_3  \right)=\varepsilon^2\left(a_3a_6+
 a_4a_5-r^\prime\chi  \right). \] In combination with the formula
 for $a_7\circ r-r^\prime a_7$ this gives \[g\circ r-r^\prime g=-3\varepsilon^2r^\prime\chi. \]
 Taking the modified KdV equation to be represented by \[f=g-3\varepsilon^2\chi,
 \]the preceding equation $g\circ r=r^\prime f$ shows that $\xymatrix{\  g & \ar[l]_r  f
 }$ as claimed.

 By Propositions 4 and 5 there exists \[s=a_1-\varepsilon a_3+\cdots\in\mathcal{A}[\varepsilon]
 \] satisfying $r\circ s=s\circ r=a_1.$ Then, by Proposition 9, \[\xymatrix{\  f & \ar[l]_s
 g,
 }\] that is, $f\circ s=s^\prime g.$

 Since both $g$ and $\chi$ are in $D\mathcal{A}\subset
 \mathcal{A},$ the same is true of all the coefficients of
 $\varepsilon^n$ in $f\circ s.$ If $s_n(a_1,a_2,\ldots)$ is the
 corresponding coefficient of $s$, then $s_n^\prime g\in
 D\mathcal{A}.$ This will show that all the $s_n$ give local conserved
 quantities for the SUSY KdV equation.
 \section{Ring Extensions and Non-Local Conservation Laws}
 In the general situation $D:\mathcal{B}\rightarrow \mathcal{B}$ of
 an oriented supercommutative ring and a superderivation, an
 element $u\in\mathcal{B}$ may or may not have the form $u=Dv$ for
 some $v\in\mathcal{B}.$ But for a fermionic $u=-\bar{u}$ one can
 always pass to the extension $ \tilde{D}:\tilde{
 \mathcal{B}}\rightarrow \tilde{ \mathcal{B}}$ where $ \tilde{
 \mathcal{B}}$ is the ring of polynomials with $ \mathcal{B}$
 coefficients in a commuting indeterminate $\lambda,$ and the new
 superderivation is $ \tilde{D}=D+u\frac{\partial}{\partial
 \lambda}.$ (If the extension was unnecesary then $ \tilde{D}v=u$
 will have more than one solution in $ \tilde{B}$.)

 The natural first example is given by $u=a_1$, the generator of $
 \mathcal{A}(a_1,a_2,\ldots),$ the free SUSY derivation ring on a
 single fermionic generator. The extension just described is $
 \mathcal{A}(a_0,a_1,a_2,\ldots),$ the free SUSY derivation ring
 on a bosonic generator $a_0,$ with $Da_n=a_{n+1}$ for $n\geq0.$

 The ring of formal power series $
 \mathcal{A}(a_0,a_1,a_2,\ldots)[\varepsilon]$ has the same
 universal property seen earlier in the fermionic case. That is,
 given any $ \tilde{D}:\tilde{
 \mathcal{B}}\rightarrow \tilde{ \mathcal{B}}$ and some formal
 power series $b=\sum_0^\infty\varepsilon^nb_n$ with all
 $b_n=\bar{b}_n\in\mathcal{B},$ the substitution operation
 $g\mapsto g\circ b$ takes
 $g=\sum_0^\infty\varepsilon^ng_n(a_0,a_1,a_2,\ldots)$ to $g\circ
 b=\sum_0^\infty\varepsilon^mg_m(b,Db,\ldots).$

 Then $\Phi(g)=g\circ b$ is a well-defined map of the power series
 ring extensions \[\Phi:\mathcal{A}(a_0,a_1,a_2,\ldots)[\varepsilon]\rightarrow\tilde{\mathcal{B}}[\varepsilon]. \]
 In fact $\Phi$ is a ring homomorphism which commutes with the
 respective involutions and satisfies $\Phi D=\tilde{D}\Phi$: The
 proof is the same as for the propositions 1 and 2 given
 earlier.

 Ring extensions of the fermionic ring $
 \mathcal{A}(a_1,a_2,\ldots)$ are now constructed so as to
 incorporate $D^{-1}$ of all the local conserved quantities of the
 SUSY KdV equation. From the formulas \[r=a_1+\varepsilon a_3-\varepsilon^2a_1a_2
 \] for the Gardner transform and \[s=s_0+\varepsilon s_1+\varepsilon^2s_2+\cdots
 \]for its inverse, which satisfy $r\circ s=s\circ r=a_1,$ one can
 compute for example
 \begin{eqnarray*} & & s_0=a_1 \\ & & s_1=-a_3 \\ & & s_2=a_5+a_1a_2 \\ & & s_3=-a_7-2a_1a_4-2a_2a_3 \\
 & & s_4=a_9+\left(3a_1a_6+3a_2a_5+5a_3a_4  \right)+2a_1a_2a_2.     \end{eqnarray*}
 It was shown before that \[ f\circ s=s^\prime g\] for
 \begin{eqnarray*} & & g=a_7+3a_1a_4+3a_2a_3 \\ & & f=g-3\varepsilon^2\left(a_1a_2a_4+a_2a_2a_3  \right).
 \end{eqnarray*} However $f=Dh$ for
 $h=(a_6+3a_2a_2-3a_1a_3)+\varepsilon^2(3a_1a_1a_3-2a_2a_2a_2).$

 Therefore $D(h\circ s)=s^\prime g.$

 As pointed out before, this is a proof that $s_0,s_1,\ldots$ are
 conserved quantities for the SUSY KdV equation.

 For each $s_n$ the ring extension is made which incorporates
 $\lambda_n=D^{-1}s_n.$ Done sucesively for $s_0,s_1,\ldots$ this
 gives $
 \tilde{\mathcal{B}}=\mathcal{A}(\lambda_0,\lambda_1,\ldots),$ the
 ring of polynomials in the commuting indeterminates
 $\lambda_0,\lambda_1,\ldots,$ with coefficients in $
 \mathcal{A}(x_1,x_2,\ldots).$ The new superderivation $ \tilde{D}:\tilde{
 \mathcal{B}}\rightarrow \tilde{ \mathcal{B}}$ is $
 \tilde{D}=D+\sum_{n=0}^\infty s_n\frac{\partial}{\partial
 \lambda_n}.$

 $\smallskip$

 Supposing $\mu(\lambda_0,\lambda_1,\ldots)$ to be a polynomial
 with constant coefficients we ask for the first variation with
 respect to $g$. When $\mu=\lambda_n$ this is
 \begin{eqnarray*}  \dot{\lambda_n} &=& \frac{d}{dt}|^{t=0}D^{-1}\left( s_n\circ\left(a_1+tg \right)
 \right) \\ & = & D^{-1}\left(s_n^\prime g \right) \\ &=&
 {\left(h\circ s
 \right)}_n\in \mathcal{A}(a_1,a_2,\ldots),
 \end{eqnarray*} where ${(h\circ s)}_n$ is the coefficient of
 $\varepsilon^n$ in the power series $h\circ s.$

 This shows that $\mu(\lambda_0,\lambda_1,\ldots)$ is a conserved
 quantity if \[\sum_{n=0}^\infty \dot{\lambda_n}\frac{\partial \mu}{\partial \lambda_n}\in\tilde{D}
 \left(\tilde{\mathcal{B}}
 \right). \]
 \begin{teorema} The coefficients of $e^{\varepsilon \lambda}$ are all nonlocal conserved quantities
 for the algebraic version of the SUSY KdV equation.  \end{teorema}
 \begin{demosteor}{\em With \begin{eqnarray*} & &
 e^{\varepsilon\lambda}=1+\varepsilon\mu_1+\varepsilon^2\mu_2+\cdots
 \\ & & \frac{\partial}{\partial \lambda_n}e^{\varepsilon
 \lambda}=\varepsilon^{n+1}e^{\varepsilon\lambda}
 \end{eqnarray*} one has $\frac{\partial}{\partial
 \lambda_n}\mu_p=\mu_{p-n-1}$ and $
 \dot{\mu_p}=\sum_{n=0}^{p-1}\mu_{p-n-1}\dot{\lambda_n}.$

 This is the coefficient of $\varepsilon^{p-1}$ in the power
 series $e^{\varepsilon\lambda}(h\circ s).$

 Evidently $e^{\varepsilon\lambda}(h\circ s)\in\tilde{B}[\varepsilon].$

 The proof of the theorem is complete when we have shown that \[e^{\varepsilon\lambda}\left(h\circ s \right)
 \in\tilde{D}\left(\tilde{B}[\varepsilon]  \right).\] However the
 substitution operation $\Phi(g)=g\circ\lambda$ gives a ring
 homomorphism \[\Phi:\mathcal{A}(a_0,a_1,\ldots)[\varepsilon]\rightarrow\tilde{B}[\varepsilon]. \]
 Obviously $\Phi(e^{\varepsilon a_0})=e^{\varepsilon\lambda},$
 while \begin{eqnarray*} & & \Phi(a_1)=a_1\circ \lambda=D\lambda=s \\ & & \Phi(a_n)=D^ns
 \hspace{2mm}\mathrm{\:for\:} \hspace{2mm}n\geq1.   \end{eqnarray*}
 This shows that \[\Phi(h)=h\circ s,\] giving
 \[\Phi\left(e^{\varepsilon a_0}h(a_0,a_1,\ldots,\varepsilon) \right)=e^{\varepsilon\lambda}\left(h\circ s \right). \]
 The search for antiderivatives can therefore be done in the more
 accessible ring $ \mathcal{A}(a_0,a_1,\ldots)[\varepsilon].$
 Indeed \[e^{\varepsilon a_0h}=Dl \] with $l=e^{\varepsilon
 a_0}(F_0+\varepsilon F_1+\varepsilon^2F_2)$ and $F_0,F_1,F_2$
 certain fermionic elements of $ \mathcal{A}(a_1,a_2,\ldots).$ The
 desired equation reduces to \begin{eqnarray*} & & DF_0=a_6+3a_2a_2-3a_1a_3 \\
 & & DF_1+a_1F_0=0 \\ & & DF_2+a_1F_1=3a_1a_2a_3-2a_2a_2a_2 \\ & & a_1F_2=0.
 \end{eqnarray*} These equations are satisfied by \[
 F_0=a_5+3a_1a_2,F_1=a_1a_4-a_2a_3,F_2=-2a_1a_2a_2.\]
 Because the ring homomorphism satisfies $\Phi D=\tilde{D}\Phi$ we
 conclude that from \[\Phi(e^{\varepsilon a_0}h)=e^{\varepsilon
 \lambda}(h\circ s)\]
we may infer \[e^{\varepsilon \lambda}(h\circ s)=\tilde{D}(\Phi
l).\]
 This completes the proof of the theorem.
 }\end{demosteor}
 \section{Conservation Laws for Superfields}
 The algebraic constructions done so far will now be applied to
 the SUSY KdV equation. This equation deals with superfields,
 which may be described as follows.

 Suppose $\Lambda$ is a finite dimensional Grassmann algebra
 generated by anticommuting elements $\theta,\eta_1,\eta_2,\ldots$
 which satisfy $\theta^2=\eta_1^2=\eta_2^2=\cdots=0.$

 Any element of $\Lambda$, after reorderings and sign changes, may
 be written uniquely as \[\phi=v(\eta_1,\eta_2,\ldots)+\theta
 u(\eta_1,\eta_2,\ldots).\] Then the superderivation
 $\frac{\partial}{\partial \theta}:\Lambda\rightarrow\Lambda$ is
 defined by $\frac{\partial\phi}{\partial \theta}=u.$

 A superfield is any infinitely differentiable function
 $\phi:\mathbb{R}\rightarrow\Lambda,$ and the ring of all
 superfields is denoted by $ \mathcal{C}^\infty(
 \mathbb{R},\Lambda).$ To avoid confusion with the algebraic case,
 the superderivation in this ring is written
 $D_1=\frac{\partial}{\partial\theta}+\theta\frac{\partial}{\partial x}.$

 Thus $\phi(x)=v(x)+\theta u(x)$ and $D_1\phi=u(x)+\theta
 v^\prime(x).$

 Ring homomorphisms from algebra to analysis are given by
 substitution of elements of $ \mathcal{C}^\infty(
 \mathbb{R},\Lambda).$ For example if $ \phi=-\bar{\phi}$ in $ \mathcal{C}^\infty(
 \mathbb{R},\Lambda)$ one has the ring homomorphism \[ \mathcal{A}\left(a_1,a_2,\ldots  \right)\rightarrow
  \mathcal{C}^\infty(
 \mathbb{R},\Lambda)  \] which sends $f(a_1,a_2,\ldots)$ to
 $f\circ\phi=f(\phi,D_1\phi,\ldots).$ This homomorphism
 interrelates the two superderivations, in the sense that \[D_1\left( f\circ \phi\right)=\left(Df \right)\circ\phi. \]
 The associativity equation $(g\circ
 f)\circ\phi=g\circ(f\circ\phi)$ continues to hold when
 $f=-\bar{f}$ in $ \mathcal{A}(a_1,a_2,\ldots)$ and
 $\phi=-\bar{\phi}$ in $\mathcal{C}^\infty(
 \mathbb{R},\Lambda),$ while $g\in\mathcal{A}(a_1,a_2,\ldots)$ is
 arbitrary. (The proof is the same as for Proposition 3).

 For the convergence of integrals one must work in subrings of $\mathcal{C}^\infty(
 \mathbb{R},\Lambda).$

 Let $\mathcal{C}_\downarrow^\infty(
 \mathbb{R},\Lambda)$ be the superfields that diminish rapidly at
 $x=\pm\infty$ together with all derivatives. When $\Phi$
 satisfies $D_1\Phi\in\mathcal{C}_\downarrow^\infty$, $\Phi$ and all its derivatives are bounded
 functions, and in particular $\frac{\partial}{\partial\theta}\Phi\in\mathcal{C}_\downarrow^\infty.$
  Thus $\psi\Phi\in\mathcal{C}_\downarrow^\infty$ when $\Psi\in\mathcal{C}_\downarrow^\infty$
  and $D_1\Phi\in\mathcal{C}_\downarrow^\infty.$

 The non-local extension of $\mathcal{C}_\downarrow^\infty$
 may be defined to be \[\mathcal{C}_{NL}^\infty( \mathbb{R},\Lambda)=
 \{\Phi\in\mathcal{C}^\infty( \mathbb{R},\Lambda):D_1\Phi\in\mathcal{C}_\downarrow^\infty(
 \mathbb{R},\Lambda)\}.  \] Then $\mathcal{C}_{NL}^\infty$ is
 again a derivation ring, and it contains $
 \mathcal{C}_\downarrow^\infty$ as an ideal. The formulas
 \begin{eqnarray*}& & \phi(x)=v(x)+\theta u(x) \\ & & D^{-1}\phi(x)=\int_{-\infty}^xu(s)ds+\theta v(x)  \end{eqnarray*}
 give an explicit mapping
 $D_1^{-1}:\mathcal{C}_\downarrow^\infty\rightarrow\mathcal{C}_{NL}^\infty,$
 with $D_1^{-1}D_1\phi=\phi$ as well as $D_1D_1^{-1}=\phi$ for all
 $\phi\in\mathcal{C}_\downarrow^\infty.$

 When $\Phi(x)=V(x)+\theta U(x)\in\mathcal{C}_{NL}^\infty$ one can
 define the integral of $\Phi(x)$ to be \[\int\Phi=\int_{-\infty}^\infty u(x)dx. \]
 Thus, integration is an additive mapping from
 $\mathcal{C}_{NL}^\infty( \mathbb{R},\Lambda)$ to the Grassmann
 algebra $\Lambda$.

 And, when
 $\phi\in\mathcal{C}_\downarrow^\infty\subset\mathcal{C}_{NL}^\infty$
 one has \[\int D_1\phi=0. \] These preparations done we turn to
 the SUSY KdV equation, which is represented by
 $g=a_7+3a_1a_4+3a_2a_3: $ if $\phi$ is a time-dependent
 superfield then \[ \frac{\partial}{\partial t}\phi=\dot{\phi}=g\circ
 \phi.\] With $s=\sum_0^\infty\varepsilon^ns_n(a_1,a_2,\ldots)$
 the inverse of Gardner transform, and any $
 \bar{\phi}=-\phi\in\mathcal{C}_\downarrow^\infty,$ we define \[ \Phi=D_1^{-1}\left(s\circ\phi \right),
 \]a formal power series with $\mathcal{C}_{NL}^\infty$
 coefficients.

 Then \[ J\left(\phi  \right)=\int e^{\varepsilon\Phi}\] will be
 shown to be a power series whose coefficients are nonlocal
 conservation laws for the SUSY KdV equation.

 To compute $
 \dot{\Phi}=\frac{d}{dt}|^{t=0}\Phi(\phi+t\dot{\phi})$ we
 recall first that \[\frac{d}{dt}|^{t=0}s\circ\left(a_1+tg  \right)=D\left( h\circ s \right)
 \]in the ring of formal power series
 $\mathcal{A}(a_1,a_2,\ldots)[\varepsilon],$ with $h=\sum_0^\infty
 h_n(a_1,a_2,\ldots)$ as computed before.

 The substitution homomorphism given by $\phi$ then gives
 $\frac{d}{dt}|^{t=0}s\circ(\phi+t\dot{\phi})=D_1(h\circ
 s\circ\phi)$.

 Consequently \begin{eqnarray*}  \dot{\Phi} & = & \frac{d}{dt}|^{t=0}D_1^{-1}\left
 ( s\circ\left(\phi+t\dot{\phi}  \right) \right) \\ & = & h\circ s\circ\phi.  \end{eqnarray*}
 Since $ \dot{J}=\varepsilon\int e^{\varepsilon\Phi}\dot{\Phi},$
 the proof will be complete when it has been shown that \[\int e^{\varepsilon\Phi}\left(h\circ s\circ \phi
  \right)=0.   \] However, it was shown earlier that there exists
  $F=\sum_0^\infty\varepsilon^nF_n(a_1,a_2,\ldots)$ satisfying
  \[e^{\varepsilon a_0}h=D(e^{\varepsilon a_0}F)\] in the ring $
  \mathcal{A}(a_0,a_1,\ldots)[\varepsilon]$ of formal power series
  with $ \mathcal{A}(a_0,a_1,\ldots)$ coefficients. Under the
  operation of substitution by $\Phi$ this equation becomes \[e^{\varepsilon\Phi}\left(h\circ s\circ\phi  \right)
  =D_1\left( e^{\varepsilon\Phi}\left(F\circ s\circ\phi  \right) \right)
  \] in the ring $\mathcal{C}_{NL}^\infty(
  \mathbb{R},\Lambda)[\varepsilon],$ because $D_1\Phi=s\circ \phi$
  and $h$ and $F$ do not involve $a_0$.

  The coefficients of $e^{\varepsilon\Phi}$ are in
  $\mathcal{C}_{NL}^\infty$ while the coefficients of $F\circ
  s\circ \phi$ are in $ \mathcal{C}_\downarrow^\infty$.

  Therefore their product is in $ \mathcal{C}_\downarrow^\infty$,
  where $ \psi\in \mathcal{C}_\downarrow^\infty$ implies $\int
  D_1\psi=0.$

  This completes the proof that $J(\phi)=\int e^{\varepsilon\Phi}$
  is a conserved quantity for the SUSY KdV equation.

  $\smallskip$

  In closing we may compare $\int e^{\varepsilon\Phi}$ with some
  conserved quantities found in the literature. Starting with
  $\phi=-\bar{\phi}\in\mathcal{C}_\downarrow^\infty,$ the first
  few coefficients of
  $\Phi=\sum_0^\infty\varepsilon^n\Phi_n=D^{-1}(s\circ\phi)$ can
  be found from the corresponding coefficients of the inverse
  Gardner transform
  $\sum_0^\infty\varepsilon^ns_n(a_1,a_2,\ldots).$ After replacing
  $D_1$ by the shorter notation $D=\frac{\partial}{\partial
  \theta}+\theta\frac{\partial}{\partial x}$ one finds that
  \begin{eqnarray*} & & \Phi_0=D^{-1}\phi \\ & & \Phi_1=-D\phi \\ & & \Phi_2=D^3\phi+D^{-1}(\phi D\phi)
  \\ & & \Phi_3=-D^5\phi-2{(D\phi)}^2+2\phi(D^2\phi).
  \end{eqnarray*}These elements of $\mathcal{C}_{NL}^\infty$ are
  all bosonic, and the first few coefficients of
  $e^{\varepsilon\Phi}=1+\sum_{n=1}^\infty\Delta_n$ are
  \begin{eqnarray*} & & \Delta_1=\Phi_0 \\ & & \Delta_2=\frac{1}{2}\Phi_0^2+\Phi_1 \\ & &
  \Delta_3=\frac{1}{6}\Phi_0^3+\Phi_0\Phi_1+\Phi_2
  \\ & & \Delta_4=\frac{1}{24}\Phi_0^4+\frac{1}{2}\Phi_0^2\Phi_1+\Phi_0\Phi_2+\frac{1}{2}\Phi_1^2+\Phi_3.
  \end{eqnarray*} Because we are only interested in the integrals
  of $\Delta_n(\phi)$, terms which fall into
  $D\mathcal{C}_\downarrow^\infty$ can be left out because they
  have identically zero integrals. For example \[\Phi_0\Phi_1=-\left(D^{-1}\phi  \right)\left( D\phi \right)
  =-D\left(\left(D^{-1}\phi  \right)\phi   \right).   \] After
  rewriting the $\Delta_n$ in terms of $D^{-1}\phi,\phi,\ldots$
  and simplifying in the manner just described we arrive at
  \begin{eqnarray*} & & \Delta_1=D^{-1}\phi \\ & & \Delta_2=\frac{1}{2}{(D^{-1}\phi)}^2 \\ & &
  \Delta_3=\frac{1}{6}{(D^{-1}\phi)}^3+D^{-1}(\phi D\phi)
  \\ & &
  \Delta_4=\frac{1}{24}{(D^{-1}\phi)}^4-\frac{1}{2}{(D\Phi)}^2+(D^{-1}\phi)D^{-1}(\phi
  D\phi).
  \end{eqnarray*} Replacing $\phi$ by $-\phi$ in these formulas we
  obtain constant multiples of the integrands which appear in the
  nonlocal conserved quantities
  $J_{_{\frac{1}{2}}},J_{_{\frac{3}{2}}},J_{_{\frac{5}{2}}},J_{_{\frac{7}{2}}}$
  presented in reference \cite{M4 and Dargis}.

  The sign change comes from the ambiguity $g=\pm
  a_7+3a_1a_4+3a_2a_3$ in the definition of the SUSY KdV equation.

  The two versions are interchanged by the transformation $T:\mathcal{A}
(a_1,a_2,\ldots)\rightarrow\mathcal{A} (a_1,a_2,\ldots)$ given by
$(Tf)(a_1,a_2,\ldots)=-f(-a_1,-a_2,\ldots).$

This transformation is not a ring homomorphism but it satisfies
$DT=TD$. In terms of the substitution operation,
$Tg=-(g\circ(-a_1))$ in general, with $Tf=(-a_1)\circ
f\circ(-a_1)$ when $\bar{f}=-f$.

The associativity and the cancellation $(-a_1)(-a_1)=a_1$ then
give \[ T(g\circ f)=(Tg)\circ(Tf). \]

Therefore $T$ also exchanges the respective Gardner transforms and
conservation laws.
\section{Conclusions}
We introduced the fermionic substitution semigroup and the
resulting Gardner category. We proved several propositions
concerning their algebraic structure. This algebraic framework
allows to define general Gardner transformations between different
non-linear SUSY differential equations. We then introduced a SUSY
ring extension which alowed to consider in the same algebraic
setting the construction of all the known non-local conserved
quantities of $N=1$ SKdV.

The algebraic version of the non-local conserved quantities was
solved in terms of the exponential function applied to the
$D^{-1}$ of the local conserved quantities of $N=1$ SKdV. Finally
the same formulas were shown to work for rapidly decreasing
superfields.

$\smallskip$

\textbf{Acknowledgment} We thank Professor Rafael Diaz (IVIC,
Venezuela) for pointing out reference \cite{Witten} to us.

\end{document}